\documentclass[]{aa}
\usepackage{graphics,latexsym}     
\begin{document}
\title{The first detection of weak gravitational shear in infrared
observations: Abell 1689}
\author{L.J. King$^{1,2}$, D.I. Clowe$^{1}$, C. Lidman$^{3}$,
        P. Schneider$^{1,2}$, T. Erben$^{1,4,5}$,        J.-P. Kneib$^{6}$, G. Meylan$^{7}$}
\institute{
1: Institut f\"{u}r Astrophysik und Extraterrestrische
   Forschung, Auf dem H{\"u}gel 71, D-53121 Bonn, Germany\\
2: Max-Planck-Institut f\"{u}r Astrophysik, 
   Karl-Schwarzschild-Strasse 1, D-85741 Garching, Germany \\
3: European Southern Observatory, Casilla 19, Santiago, Chile\\
4: Institut d'Astrophysique de Paris, 98 bis Boulevard Arago, 
   F-75014 Paris, France\\
5: Observatoire de Paris, DEMIRM, 61 Avenue de l'Observatoire, 
   F-75014 Paris, France\\
6: Observatoire Midi-Pyr\'en\'ees, 14 Av. E. Belin, 
   F-31400 Toulouse, France\\
7: Space Telescope Science Institute, 3700 San Martin Drive, 
   Baltimore, MD 21218, USA\\  
}
\date{}
\authorrunning{L.J. King et al.}
\titlerunning{Gravitational shear from an infrared study of Abell 1689}
\abstract{
We present the first detection of weak gravitational shear at infrared
wavelengths,  using observations  of the  lensing cluster  Abell 1689,
taken with  the SofI camera on  the ESO-NTT telescope.  The imprint of
cluster lenses on  the shapes of the background  galaxy population has
previously   been   harnessed  at   optical   wavelengths,  and   this
gravitational shear  signal enables  cluster mass distributions  to be
probed, independent  of whether  the matter is  luminous or  dark.  At
near-infrared wavelengths, the  spectrophotometric  properties of  galaxies
facilitate  a clean  selection of  background objects  for use  in the
lensing analysis.  A  finite-field mass reconstruction and application
of  the aperture  mass  ($M_{\rm ap}$)  statistic  are presented.  The
probability   that   the   peak   of  the   $M_{\rm   ap}$   detection
($\frac{S}{N}\sim 5$),  arises from  a chance alignment  of background
sources is  only $\sim 4.5\times 10^{-7}$. The  velocity dispersion of
the  best-fit singular  isothermal  sphere model  for  the cluster  is
$\sigma_{\rm  1D}=1030^{+70}_{-80}\,{\rm km}\,{{\rm s}^{-1}}$,  and we
find  a K-band mass-to-light  ratio of  $\sim 40\,M_{\odot}/L_{\odot}$
inside a $0.44\,{\rm Mpc}$ radius.
}
\def\A{{\cal A}}
\def\eck#1{\left\lbrack #1 \right\rbrack}
\def\eckk#1{\bigl[ #1 \bigr]}
\def\rund#1{\left( #1 \right)}
\def\abs#1{\left\vert #1 \right\vert}
\def\wave#1{\left\lbrace #1 \right\rbrace}
\def\ave#1{\left\langle #1 \right\rangle}
\def\arcsecf {\hbox{$.\!\!^{\prime\prime}$}}
\def\arcminf {\hbox{$.\!\!^{\prime}$}}
\def\bet#1{\left\vert #1 \right\vert}
\def\vp{\varphi}
\def\vt{{\vartheta}}
\def\map{{$M_{\rm ap}$}}
\def\d{{\rm d}}
\def\mj{$\rm {m_{j}}$}
\def\mj{$\rm {m_{k}}$}
\def\col{$\rm {m_{j}}-\rm {m_{k}}$}\def\eps{{\epsilon}}
\def\vc{\vec} 
\def\s{{\rm d}}
\def\s{{\rm s}}
\def\t{{\rm t}}
\def\E{{\rm E}}
\def\L{{\cal L}}
\def\i{{\rm i}}

{\catcode`\@=11
\gdef\SchlangeUnter#1#2{\lower2pt\vbox{\baselineskip 0pt \lineskip0pt
\ialign{$\m@th#1\hfil##\hfil$\crcr#2\crcr\sim\crcr}}}}

\def\gtrsim{\mathrel{\mathpalette\SchlangeUnter>}}
\def\lesssim{\mathrel{\mathpalette\SchlangeUnter<}}      

\maketitle
\section{Introduction}
A cluster acting  as a weak gravitational lens  distorts the shapes of
background galaxies  by virtue of its tidal  gravitational field.  The
seminal paper  of Kaiser \& Squires  (1993) describes how  to use this
information  to  obtain  a  parameter-free reconstruction  of  a  mass
distribution.  Their technique, and variants thereof, has been applied
to  optical  observations of  a  number  of  clusters (e.g.  Clowe  et
al. 2000; Hoekstra et al. 2000).

Prior  to  this  work, there  has  been  no  published report  on  the
detection of gravitational shear in infrared observations. For a given
integration time, the attainable number density of background galaxies
for use in weak lensing studies is usually much lower than for optical
observations. In addition, infrared detectors currently lag behind 
optical detectors in their field-of-view making them even less
efficient for wide-field studies. Nevertheless, with  developments in detector technology,
and  new wide-field infrared  cameras becoming  available, this  is an
avenue  to  be  explored.  Gray  et al.  (2000)  have  considered  the
magnification   induced  depletion   effect  in   near-infrared  CIRSI
observations of the cluster Abell 2219, in order to obtain 
best-fit SIS (singular isothermal sphere) and NFW (Navarro, Frenk \&
White 1996) models.

Here we report on J, H and Ks infrared observations in which we detect
the  gravitational   shear  signature   of  the  cluster   Abell  1689
($z=0.182$),   manifest  in   the  distorted   images   of  background
galaxies. We start by outlining the observations, and the strategy for
obtaining  a catalogue  of background  galaxies  for use  in the  weak
lensing  analysis. Then  we describe  how this  catalogue was  used to
perform a  mass reconstruction of the cluster.  The \map$\,$ statistic
of  Schneider (1996)  is then  calculated, along  with  a quantitative
measure of its  significance. Next we use Ks  selected cluster members
to determine the luminosity of  the lens, and obtain its mass-to-light
ratio  by  comparison  with  the  mass derived  from  fitting  an  SIS
 profile  to the  weak lensing  data.  We
conclude with a brief discussion  of the results, and the implications
for  future work. For  quantitative estimates,  we assume  a cosmology
with   $\Omega=0.3$,  $\Lambda=0.7$   and   $H_{0}=70\,{\rm  km}\,{\rm
s}^{-1}\,{\rm Mpc}^{-1}$, and that the background galaxy population is
at  $z=1.0$.  The conversion  between  angular  and  linear scales  at
$z=0.182$ is $1\arcminf 0 \equiv 0.18\,{\rm Mpc}$.

\section{Observations and data reduction}
Our data were  obtained on the nights 2000 April  29th and 30th, using
the SofI  camera (Moorwood  et al. 1998)  on the ESO-NTT  telescope at
Cerro La Silla Observatory, Chile.  The instrument was used in ``Large
Field Imaging"  mode, where the  field of view  is $4\arcminf 9$  on a
side, and  the pixel  scale is $0\arcsecf  29$/pixel.  Abell  1689 was
observed in three filters, J,  H and Ks. 
Individual integrations  lasted between 10 and $30\,{\rm
s}$  and  several  of  these  were  co-added  to  form  an  individual
image. Between images, the telescope was moved by up to $2\arcminf 0$,
while always keeping the target  within the field of view. This offset
is somewhat larger  than is typical in such  observations; however, it
was  necessary given the  high density  of galaxies  in the  centre of
A1689.  The field of view  is effectively increased to $\sim 7\arcminf
0$ (1440  pixels).  During the  observations the seeing  was typically
$\sim 0\arcsecf  8$ and conditions were  perfectly photometric, except
during the early hours of the first night.

The  data were  reduced  in the  standard  way. From  each image,  the
zero-level offset was removed, {the flat field correction was applied,} and then an estimate
of  the sky from  other images  in the  sequence was  subtracted.  The
images were combined so that the $\frac{S}{N}$ (signal-to-noise) ratio
of the final image was maximised {and with linear registration relative 
to the first frame in each sequence}. 

The photometric  calibration was done by observing  several stars from
the   list   of  HST   NICMOS   photometric   standards  (Persson   et
al.  1998). Although there  was some  thin cirrus  early in  the first
night,  we  were  able  to  observe a  sufficiently  large  number  of
standards to derive accurate  instrumental zero points and atmospheric
extinction  coefficients in  each  of the  three  filters.  The  total
integration times ($t$), the final image quality (IQ), the sensitivity
limit  (1-$\sigma$  noise  limit  in  1 square  arcsecond)  ($SL$)  and
extinction coefficients (A) of the combined  images in the J, H and Ks
bands are  given in  Table 1. The  instrumental zero points  were very
stable  over  the  two nights  and  the  scatter  in  the fit  of  the
instrumental magnitude versus airmass  is less than 0.01 magnitudes in
all three filters.

We  did not  correct for  colour terms  between the  SofI instrumental
system  and  that  used  for  the HST  NICMOS  standards.  Preliminary
estimates   suggest  that  these   terms  are   all  less   than  0.02
magnitude. Nor did we correct the flat fields for illumination effects
which cause the photometry to  vary differentially across the field by
up to 0.03 magnitudes. Thus, the accuracy of the relative and absolute
photometry is around 0.03 magnitudes for both.

\begin{table}
\caption{The integration  time ($t$), image  quality (IQ), sensitivity
limit ($SL$) and extinction (A) are shown for each of the filters with
which A1689 was observed.}
\begin{tabular}{l c c c c}
\hline
Filter&$t$ (ks)&IQ (")&$SL$ (mag)&A (mag/airmass)\\
\hline
J&14.4&0.77&25.59&0.07\\
H&10.8&0.85&24.21&0.04\\
Ks&15.6&0.73&23.57&0.07\\
\hline
\end{tabular}
\end{table}

\section{Obtaining a catalogue of background objects}

The goal described  in this section is the  acquisition of a catalogue
of background object positions and  ellipticities, that can be used in
the weak lensing analysis.
 
Our strategy was to start by using SExtractor (Bertin \& Arnouts 1996)
to detect objects  on each of the J, H and  Ks images separately.  For
all filters, an object detection  and analysis threshold of at least 3
contiguous pixels  above $1 \sigma$  was required.  Since the  J image
contained most detections, the H and Ks images were transformed to the
J reference frame, using IRAF tasks (see Tody 1993 and
references therein). SExtractor was then run
in  ``double-image mode",  using the  J image  for detection  (with an
object  detection and  analysis  threshold of  at  least 2  contiguous
pixels above  $0.7 \sigma$)  and each of  the J,  H and Ks  images for
measurements. For each object, magnitudes were measured using both a 15
pixel ($4\arcsecf4$) diameter aperture, and an outer isophote at 70\%
of the sky noise (J: 24.07, H: 23.11, Ks: 22.55 mag/sq.\,arcsec.). 
Aperture magnitudes are used for colour determination, and isophotal magnitudes are used for all other purposes.
The  LDAC   tools\footnote{Freeware   available  from
ftp://ftp.strw.leidenuniv.nl/pub/ldac/} were used to
obtain a single catalogue of these sources.

Ellipticities for the objects  were determined from the second moments
of their surface brightness using  a modified version of Nick Kaiser's
IMCAT\footnote{http://www.ifa.hawaii.edu/$\sim$kaiser/imcat} software. Bright
(but unsaturated) stars were selected on the basis of their half-light
radii and were used to determine  the point spread function (PSF) as a
function of position  in the images. A bi-cubic  polynomial was fit to
the stellar ellipticities {(for 25 stars)} and used to correct the ellipticities of the
galaxies using  the technique  detailed in Kaiser  et al.  (1995). The
smearing  of the  galaxy ellipticities  by  the now  circular PSF  was
removed using the basic method  outlined in Luppino and Kaiser (1997),
but where the pre-seeing shear  polarizability tensor was fit as a 5th
order polynomial  function of $r_g$, the Gaussian  smoothing radius at
which the  galaxy achieves  maximum significance from  the background,
and the post-seeing ellipticity. The resulting shear estimate has been
shown to be good to a  few percent accuracy for well behaved PSFs such
as those in these images (Erben et al.~2001).

Stellar  objects  were  filtered  out,  based  on  their  locus  in  a
magnitude-radius  plot. Next, making  use of  the small  dispersion in
colour at a particular  magnitude for cluster galaxies, likely cluster
members  were excised to  exclude even  the faintest  cluster members,
where the dispersion in the cluster sequence becomes larger.

Further, only galaxies fainter  than ${\rm m_{\rm J}}=16.6$ and having
a  $\frac{S}{N} >  5$  in  J were  retained.  The $\frac{S}{N}$  limit
corresponds to  m$_{\rm J}=23.2$, for the smallest  objects of similar
size  to the  PSF.  Finally, only galaxies  redder  than the  cluster
sequence  galaxies  were kept  in  the  catalogue (they  are  most
probably at  higher redshifts  than the cluster; see for  example the
colour evolution plots in Poggianti 1997).

\begin{figure}
\resizebox{8cm}{!}{\includegraphics{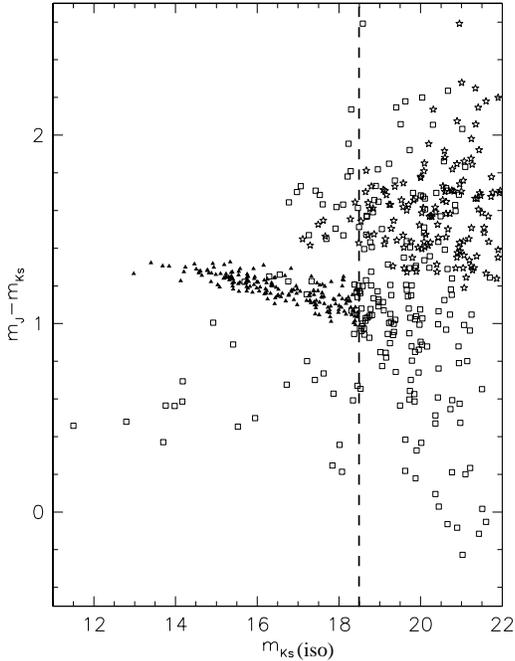}}
\hfill
\caption{The   colour-magnitude  diagram   for  objects   selected  as
background galaxies for use in  the lensing analysis (open stars). The
dashed line  at m$_{\rm  Ks}=18.5$ indicates the  limit down  to which
objects  selected as  cluster  galaxies were  used  in the  luminosity
estimation (solid triangles).  Other  objects detected but not used in
the  lensing or  mass-to-light analysis,  including stars, fainter
cluster members, {and cluster members outside the radius
inside which the mass-to-light ratio is determined} are marked with open squares. Here, m$_{\rm Ks}$ and
m$_{\rm   J}$  are aperture magnitudes, and m$_{\rm Ks}$(iso) is an 
isophotal magnitude.  Note the clear colour-magnitude sequence of cluster galaxies.}
\label{cmd}
\end{figure}

The selection process for the  catalogue as outlined above is the most
conservative of  our trials, and the  final number of  sources used in
the  weak  lensing analysis  is  157,  just  over 3/arcmin$^{2}$.   In
Fig.\,\ref{cmd},  these  objects  are  indicated  (open  stars)  on  a
colour-magnitude diagram  for the observations.   We experimented with
various cuts,  which induce only a  marginal change in  the results of
the lensing analysis described in the next Section.

\section{Weak Lensing Analysis}
For the weak lensing analysis, we used the mean of the shear estimates
determined  from the  separate J,  H and  Ks images,  weighted  by the
$\frac{S}{N}$ of the object detections.  The best-fit SIS model to the
radial shear profile  (centred on the brightest cluster  galaxy) has a
velocity dispersion $\sigma_{\rm 1D}=1030^{+70}_{-80}\,{\rm km}\,{{\rm
s}^{-1}}$,  with a  significance of  6.1  relative to  the null  model
($\sigma_{\rm  1D}=0\,{\rm   km}\,{{\rm  s}^{-1}}$).  This   value  is
consistent  with  the  best-fit  SIS  to  WFI optical  data  ($\sigma_{\rm
1D}=998^{+33}_{-42}\,{\rm km}\,{{\rm  s}^{-1}}$) presented in  King et
al. (2001). {The best-fit NFW model has $r_{200}=1.94\,{\rm Mpc}$
and $c=5.7$, the significance and quality of fit being the same as the 
SIS.} We checked  that fitting a  radial shear profile  to the
objects bluer  than the cluster  sequence results in  an insignificant
weak lensing signal.

\subsection{Mass reconstruction}
\begin{figure}
\resizebox{8.5cm}{!}{\rotatebox{-90}{\includegraphics{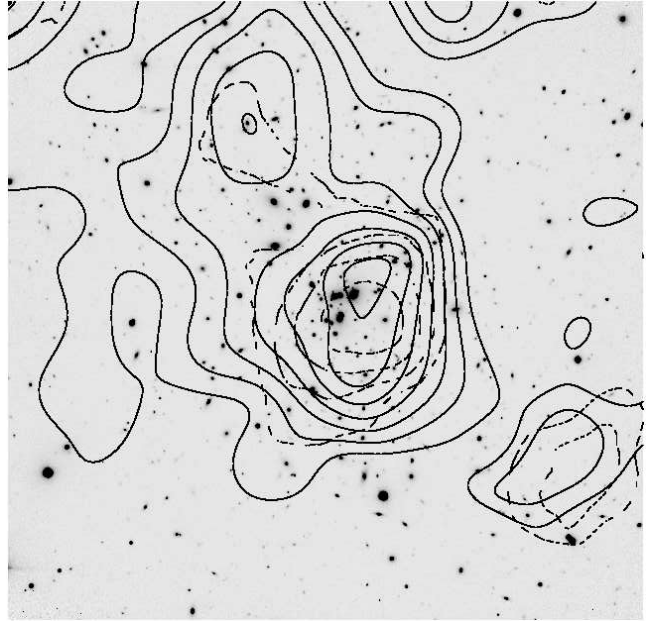}}}
\hfill

\caption{ The greyscale shows the  SofI J-band image, $7\arcminf 0$ on
a side, N to the top and E to the left.   The solid  contours show the  surface mass  density $\kappa$
from  a finite-field  mass reconstruction  with a  smoothing  scale of
$0\arcminf 54$, plotted  at intervals of 0.05 in  $\kappa$ relative to
the  boundary value.   A  contour plot  of  $\frac{S}{N}$ for \map$\,$ with  a filter scale of $\eta=1\arcminf 95$
is     plotted     as     dashed    contours,     corresponding     to
$\frac{S}{N}=1,2,3,4$. }
\label{rec}
\end{figure}

We use  the non-parametric finite-field  mass reconstruction algorithm
detailed  in  Seitz  \&   Schneider  (2001).  The  reconstruction  was
performed with  a smoothing  scale of $0\arcminf  54$.  The  result is
shown  as solid contours  in Fig.\,\ref{rec},  normalised (necessary because of the mass-sheet degeneracy) so  that the
mean  surface  mass  density  on  a $0\arcminf  097$  wide  border  is
$\bar\kappa =0.1$. Note that the  mass peak is consistent with that of
the luminous infrared image.

Since the noise properties  for non-parametric reconstructions are not
easily  interpreted, multiple  reconstructions  were performed  during
which the catalogue object positions were kept fixed and ellipticities
were  assigned  at  random  (with replacement)  from  the  ellipticity
ensemble. This enabled us to get a handle on the noise properties, and
confirms that the central mass concentration is not a noise artefact.

\subsection{Aperture mass statistic}
Next we  consider the \map$\,$  statistic, which measures  the surface
mass density convolved with  a compensated Mexican Hat filter function
${\cal U}$.  Equivalently, the suitably weighted  ellipticities give a
practical estimator of \map
\begin{equation}
M_{\rm ap}(\vc\vt)=\frac{\pi\eta^{2}}{N}\sum_{i}\epsilon_{{\rm
t}i}(\vc\vt){\cal Q}(|\vc\theta_{i}-\vc\vt|),
\end{equation}
where  the  sum  extends  over  $N$  galaxies,  located  at  positions
$\vc\theta_{i}$,  inside  a   (filter)  radius  $\eta$ centred on position
$\vc\vt$.  The  quantity $\epsilon_{\rm t}(\vc\vt)$  is the tangential
component of  the ellipticity with  respect to position  $\vc\vt$, and
${\cal Q}$  is a filter function  related to $\,{\cal U}$.  We use the
simplest form of the filter functions from Schneider et al. (1998).

An  asset of  this statistic  is that  the noise  properties  are well
understood, and  the significance of  a \map$\,$ detection  is readily
calculated by comparing its value with that obtained from the same set
of galaxies  with random  position angles. The  signal to  noise ratio
$\frac{S}{N}$ for \map$(\vc\vt)$ is $\frac{M_{\rm ap}}{\sqrt{\sigma^{2}}}(\vc\vt)$ where $\sigma^{2}=
\ave{M_{\rm ap}^{2}}_{\rm randomisations}(\vc\vt)$.
The  $rms$ dispersion  of \map$\,$  is  calculated in  the absence  of
lensing, since the deviation from  this approximation is very small in
the weak lensing regime {; it gives an upper limit on $\sigma$}.  The $\frac{S}{N}$ of \map\, calculated 
for a  filter scale  $\eta=1\arcminf 95$  and  for 1\,000
randomisations, is  shown as  dashed contours in
Fig.\,\ref{rec}. $\frac{S}{N}\sim 5$  for the  peak of the  detection; its  location is
consistent with the peak of the non-parametric mass reconstruction and
with the brightest  cluster galaxy on the infrared  image.  Also, note
that the \map$\,$ detection is highly significant over a large area.

The calculation  of $\frac{S}{N}$ in the previous  paragraph relies on
the fact that \map(\vc\vt) has a Gaussian probability distribution. In
addition, \map$\,$ was calculated around the peak of the detection for
$4\times 10^{7}$ randomisations of the background galaxy orientations,
and  a higher $\frac{S}{N}$  peak was  obtained for  only 18  of these
cases.  The probability that  such a  random distribution  of galaxies
would give  a $\frac{S}{N}$  in excess of  the observed peak  value is
therefore $\sim  4.5\times 10^{-7}$.   As another check,  the position
angles of  the background galaxies were rotated  through $45^{\rm o}$, and
no highly significant \map$\,$ detection was obtained.

\section{Mass-to-light ratio}
Selection of cluster  members was performed in the  Ks-band data, down
to  isophotal m$_{\rm  Ks}=18.5$  and  using their  tight  
colour-magnitude  and
colour-colour loci. Further, we restricted selection to objects within
500 pixels ($2\arcminf 43$)  of the brightest  cluster galaxy. Outside
of this radius, the sky noise increases dramatically due to dithering
causing a decrease in exposure time. We  indicate the  objects selected  as cluster
members on Fig.\,\ref{cmd}. The  apparent magnitudes of these galaxies
were   converted   to  absolute   magnitudes,   and  subsequently   to
luminosities  using the evolutionary  and k-corrections  obtained from
the  single  burst,  solar  metallicity, Scalo  (1986) IMF
spectral energy  distribution contained  in HyperZ (Bolzonella  et al.
2000) (amounting to $-0.53\,{\rm  mag}$) and the extinction correction
from our data ($0.09\,{\rm mag}$).

\begin{figure}
\resizebox{8.5cm}{!}{\includegraphics{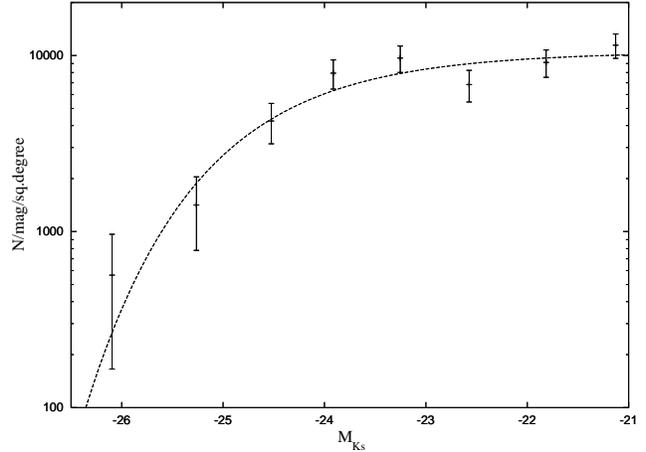}}
\hfill
\caption{The luminosity function of  the Ks selected cluster galaxies,
with  the   best-fit  Schechter  function  indicated   by  the  dashed
line. Error bars correspond to Poisson noise in each bin.}
\label{schlum}
\end{figure}

The differential K-band luminosity  function of the cluster members is
shown  in Fig.\,\ref{schlum},  with  error bars  corresponding to  the
Poisson variance in each bin, along with the best-fit Schechter (1976)
luminosity function

\begin{equation}
\Phi(L)dL=\Phi^{*}\left(\frac{L}{L_{*}}\right)^{\alpha}{\rm
exp}\left(-\frac{L}{L_{*}}\right)\frac{dL}{L_{*}}.
\end{equation}
The nonlinear least-squares  Marquardt-Levenberg algorithm was used to
obtain the  best-fit parameters  for the faint  slope $\alpha=-1.01\pm
0.14$  and  characteristic  absolute magnitude  ${\rm  M}_{*}=-24.7\pm
0.40$ (i.e. ${\rm m}_{*}=14.6$), with a reduced $\chi^{2}$ of 1.3.
Taking into  account the difference in adopted
cosmology,  this  value of  ${\rm  M}_{*}$  is  very similar  to  that
obtained by  Barger et al. (1996),  for a sample  of distant clusters.

The integrated  luminosity (extrapolated  to zero luminosity)  for the
best-fit Schechter function  is 8.0$\times 10^{12}\,L_{\odot}$ and the
direct  sum  of  galaxy   luminosities  (down  to  $L_{\rm  min}$)  is
8.8$\times 10^{12}\,L_{\odot}$. These are consistent, given the errors
of the  fit parameters.   We use the  direct sum in  the mass-to-light
ratio estimate, since we reach  well down the luminosity function: the
fraction       of      light      observed       is      approximately
$\Gamma(\alpha+2,\frac{L_{\rm        min}}{L_{*}})/\Gamma(\alpha+2)\sim
97\%$.   In order  to  estimate  the error  on  the luminosity,  first
consider   the  error   arising   from  the   various   cuts  in   the
colour-magnitude and colour-colour planes,  given that each object has
a  photometric  uncertainty.   We  started  with  the  original  uncut
catalogue, and created 100\,000  mock catalogues, assigning the $i^{\rm
th}$ object a flux  F$_{i}$=F$_{i(\rm mea)}$+ $g{\rm F}_{i(\rm err)}$,
where $g$ is a uniform gaussian random variable, and F$_{i(\rm mea)}$,
F$_{i(\rm   err)}$   are   the    measured   flux   and   its   error,
respectively. The magnitude and colour cuts were then applied to these
catalogues,  and   the  dispersion  in   the  recalculated  luminosity
determined to be only $\sim  0.35\%$ of the total.  Another very small
error  is the Poissonian  error on  the total  flux, which  amounts to
$\sim 0.05\%$. The  dominant error is the 2\%  error on the luminosity
because of  uncertainty in the Ks  zero point.  Hence  for the cluster
luminosity we take $8.8\pm  0.18\times 10^{12}\,L_{\odot}$ (which is a
lower limit  because $\sim  3\%$ of the  luminosity is expected  to be
below the Ks magnitude limit, and light outside the isophotal radii of galaxies is not accounted for).

The projected mass of the best-fit SIS to the infrared data within the
same   $2\arcminf   43$   radius  is   $M$=3.44$^{+0.49}_{-0.51}\times
10^{14}\,M_{\odot}$.  This  gives an upper-limit  K-band mass-to-light
ratio of $39.1\pm 5.7\,M_{\odot}/L_{\odot}$. {The best-fit NFW
model has a mass at this radius which is $\sim$10\% higher.}

\section{Discussion and conclusions}
We  find that  it  is possible  to  fit a  radial  shear profile  (the
best-fit   SIS  model   has   $\sigma_{\rm  1D}=1030^{+70}_{-80}\,{\rm
km\,s}^{-1}$), perform  a mass  reconstruction, and obtain  a \map$\,$
$\frac{S}{N}\sim  5$  detection using  only  the  infrared  data on  a
relatively  small   field.   The  locations   of  the  peaks   of  the
reconstruction and of the  \map$\,$ detection are consistent with that
of the brightest cluster galaxy in the infrared image.

The  K-band  luminosity  function  of  the cluster  galaxies  is  well
described   by  a  Schechter   function,  with   a  faint   end  slope
$\alpha=-1.01\pm  0.14$ and  characteristic  absolute magnitude  ${\rm
M}_{*}=-24.7\pm 0.40$. In  addition,   we  find  a   K-band  mass-to-light  ratio   of  $\sim
40\,M_{\odot}/L_{\odot}$ in the cosmology  adopted, in accord with the
study of rich Abell clusters made by Uson \& Boughn (1991).

The capability to perform  routine wide-field infrared imaging is just
around  the corner  with projects  such as the UKIRT WFCAM, the CFHT 
WIRCAM and VISTA. Weak gravitational lensing  observations  of
distorted  galaxies in  the  infrared  will
provide another window through which to probe the mass distribution on
cluster  scales.  Although  the number  density of  background sources
achievable  is  roughly  an  order  of  magnitude  lower  at  infrared
wavelengths, there are a few points worth noting:

\begin{itemize}
\item{ Colours of galaxies in the near-infrared are dominated by light
from old  evolved stars, so the  well defined cluster  sequence in the
colour-magnitude  diagram contains  both ellipticals  and  (unlike the
optical  cluster  sequence) spirals.   This  makes excising  potential
cluster members easier using infrared colours, as anything redder than
the sequence is likely to be  at a higher redshift, and anything bluer
is probably at a lower  redshift. The infrared selection of background
objects  can  then be  used  as a  consistency  check  of the  optical
selection,  to  assess  the  contamination from  cluster  members  and
foreground objects.}
\item{ With optical observations,  measurement of the ellipticities of
high redshift galaxies  can be hampered by the  presence of blue knots
of star formation.}
\item{  The total  K-band luminosity  of  the cluster  members can  be
compared with a weak lensing  mass, in order to obtain a mass-to-light
ratio. Further,  Broadhurst et al. (1992) underlined that the K-band
luminosity  is   a  good  measure  of  the   underlying  stellar  mass
irrespective of how it assembled. If a high enough resolution mass map
became  available, we  could investigate  the correlation  between any
peaks in the total mass and stellar mass content.}
\end{itemize}

Although {infrared observations are not preferable to optical ones
if the goal is simply to do a lensing analysis}, this work
highlights the complementary nature of this passband. Our next step is 
to combine the
infrared and optical information. On a galaxy-by-galaxy basis, the
selection of background galaxies can be examined. In addition, we can
compare the ellipticities of the lensed galaxies determined from each
data set. Finally, we will consider constraints from strong lensing
features to obtain the most precise description of the mass
distribution of this spectacular lens.

\begin{acknowledgements}
This work was supported by the TMR Network
``Gravitational Lensing: New Constraints on Cosmology and the
Distribution of Dark Matter'' of the EC under contract
No. ERBFMRX-CT97-0172, and by the DFG and the CNRS.
Thanks to Joan-Marc Miralles, Roser Pell\'{o}, Tom Broadhurst, 
Michael Hilker, Martin Altmann and Neil Trentham for useful discussions,
and to Leonardo Vanzi for providing us with flat field 
images from the SofI calibration plan. Thanks also to the referee for 
very helpful comments on the manuscript.
\end{acknowledgements}
  
\def\ref#1{\bibitem[1998]{}#1}

\end{document}